\documentclass[aps,prd,twocolumn]{revtex4}

\usepackage{amsfonts}
\usepackage{amsmath}
\usepackage{amssymb}
\usepackage{bm}
\usepackage{epsfig}
\usepackage{graphicx}
\usepackage{graphics}
\usepackage[usenames]{color}

\newcommand\be{\begin{equation}}
\newcommand\ba{\begin{eqnarray}}
\newcommand\ee{\end{equation}}
\newcommand\ea{\end{eqnarray}}

\newcommand\bw{\begin{widetext}}
\newcommand\ew{\end{widetext}}

\newcommand\mrm{\mathrm}

\newcommand{\nn}{\nonumber}

\newcommand{\et}{{\it et al.}}

\begin{document}
\title{A new constraint on scalar Gauss-Bonnet gravity \\ and a possible explanation for the excess of the orbital decay rate \\ in a low-mass X-ray binary}

\author{Kent Yagi}
\affiliation{Department of Physics, Montana State University, Bozeman, MT 59717, USA.}
\email{kyagi@physics.montana.edu}



\date{\today}

\begin{abstract} 

It was recently shown that a black hole (BH) is the only compact object that can acquire a scalar charge in scalar Gauss-Bonnet (sGB) theory under the small coupling approximation.
This leads to the fact that scalar radiation is emitted from a binary containing at least one BH.
In this letter, we find the constraints on this theory from BH low-mass X-ray binaries (BH-LMXBs).
The main result of this letter is that from the orbital decay rate of A0620-00, we obtained a conservative bound that is six orders of magnitude stronger than the solar system bound.
In addition to this, we look at XTE J1118+480, whose orbital decay rate has been recently measured with an excess compared to the theoretical prediction in GR due to the radiation reaction.
The cause of this excess is currently unknown.
Although it is likely that the cause is of astrophysical origin, here we investigate the possibility of explaining this excess with the additional scalar radiation in sGB theory.
We find that there still remains a parameter range where the excess can be explained while also satisfying the constraint obtained from A0620-00.
The interesting point is that for most of other alternative theories of gravity, it seems difficult to explain this excess with the additional radiation.
This is because it would be difficult to evade the constraints from binary pulsars or they have already been constrained rather strongly from other observations such as solar system experiments.
We propose several ways to determine whether the excess is caused by the scalar  radiation in sGB gravity including future gravitational wave observations with space-borne interferometers, which can give a constraint three orders of magnitude stronger than that from A0620-00.

\end{abstract}



\maketitle


\textit{Introduction:}
Testing gravitational theories~\cite{will-living} is important from both theoretical and phenomenological points of view.
For the former case, if we assume that the classical gravitational theory appears at the low-energy limit of a more fundamental theory such as superstring theory~\cite{polchinski1,polchinski2},  the theory does not necessarily reduce to general relativity (GR).
One possibility of an effective gravitational theory is Einstein-Dilaton-Gauss-Bonnet (EDGB) theory~\cite{Moura:2006pz}, where at the level of the action, the dilaton is coupled to the Gauss-Bonnet invariant with a coupling constant $\alpha$.
For the latter case, modification of gravity may naturally solve problems in GR such as dark energy, dark matter and inflation.
(See e.g. Ref.~\cite{fR}.)
So far, GR has been tested mainly in the solar system and  in binary systems, especially binary pulsars~\cite{will-living}.
We have recently shown~\cite{quadratic} that only a black hole (BH) can acquire a scalar charge in scalar Gauss-Bonnet (sGB) gravity~\cite{nojiri} under the small coupling approximation.
Hence there exists scalar dipole radiation in a compact binary system where at least one of the constituents is a BH. 
(Here, sGB gravity refers to a theory where, at the level of the action, the Gauss-Bonnet invariant is coupled to an arbitrary function of a scalar field.
EDGB theory is one specific theory in sGB gravity.)
This means that we would not be able to test this theory with a neutron-star (NS)/NS or a white-dwarf (WD)/NS binary.
In this letter, we probe this theory using BH low-mass X-ray binaries (BH-LMXBs).

One of the current bounds on sGB theory has been obtained using the Saturn probe Cassini~\cite{cassini}, which measured the Shapiro time delay, giving the constraint $\sqrt{|\alpha|} < 8.9 \times 10^{11}$cm.
Especially, for EDGB theory, a stronger constraint has been obtained from the existence of a stellar-mass BH~\cite{paniread}.
For example, the LMXB GRO J0422+32 is likely to contain a primary BH with mass $(4\pm 1)M_\odot$~\cite{gelinoGRO}.
Since a BH can only exist if $\sqrt{|\alpha|}$ is below the upper bound that is proportional to the Schwarzschild radius of a BH~\cite{kanti}, we get the 2-$\sigma$ constraint $\sqrt{|\alpha|} < 3.1 \times 10^5$cm.

LMXBs have been used to place constraints~\cite{zaglauer,psaltisBD} on several theories such as Brans-Dicke (BD) theory~\cite{brans}.
Especially, BH-LMXBs have been recently exploited~\cite{johannsen1,johannsen2} to put bounds on the size of the extra dimension in the Randall-Sundrum II braneworld model~\cite{randall2}.
We first use the orbital decay rate of A0620-00 to obtain a new constraint on sGB gravity and find that it is more than six orders of magnitude stronger than the current bound from solar system experiments~\cite{amendola}.
Among BH-LMXBs, XTE J1118+480 is an extremely interesting source since, very recently, Gonzalez \et~\cite{gonzalez} reported that there is an excess in its orbital decay rate compared to that predicted by GR. 
According to their arguments, this excess cannot be explained by the relativistic periastron precession, relativistic jets and the presence of a circumbinary disk.
In order to interpret the cause as the magnetic braking, it is required that either the mass accretion onto the BH is almost zero (which is rather implausible) or that the magnetic field of the secondary be more than 1--2 orders of magnitude greater than that typical in highly-rotating low-mass stars~\cite{gonzalez}.
Although it is most likely that this excess comes from an astrophysical origin, here, we investigate whether it can also be explained by the modification of gravity.
It would be difficult for most alternative theories of gravity since they are already constrained rather strongly, especially from binary pulsar observations.
However, sGB gravity cannot be constrained from these observations.
Indeed, we find that the possibility that the excess may be caused by the additional scalar radiation in sGB gravity has not been ruled out yet.
We propose several ways to distinguish whether the excess is caused by this additional radiation.
We use the geometrical unit $c=G=1$ throughout the letter.

\textit{sGB gravity:}
Let us consider the following action for sGB gravity~\cite{nojiri}:
\ba
S &=& \int d^4x \sqrt{-g} \bigg\{ \kappa R + \alpha f(\phi) R^2_\mrm{GB} \nn \\
& &  
-\frac{1}{2} \nabla_\mu \phi \nabla^\mu \phi + \mathcal{L}_\mrm{mat} \bigg\}\,. 
\label{action}
\ea
Here, $\kappa \equiv (16 \pi)^{-1}$, $g$ stands for the determinant of the metric $g_{\mu\nu}$ and $R$ is the Ricci scalar.
$R^2_\mrm{GB}$ is the Gauss-Bonnet invariant defined as $R^2_\mrm{GB} \equiv R^2 - 4 R_{\mu\nu}R^{\mu\nu} +R_{\mu\nu\rho\sigma}  R^{\mu\nu\rho\sigma}$ with $R_{\mu\nu}$ and $R_{\mu\nu\rho\sigma}$ representing the Ricci and Riemann tensors.
$\alpha$ is the coupling constant of the theory while $\mathcal{L}_\mrm{mat}$ corresponds to the matter Lagrangian density.
 $f(\phi)$ is an arbitrary function of the scalar field $\phi$ and $f(\phi) = e^{\phi}$ represents EDGB theory~\cite{Moura:2006pz,Pani:2009wy}.
 In $c=G=1$ units, the scalar field $\phi$ is dimensionless while the coupling constant $\alpha$ has a unit of $(\mrm{length})^2$.
Following Ref.~\cite{quadratic}, we expand $f(\phi)$ around the asymptotic value $\phi =0$ at spatial infinity as $f(\phi) = f(0) + f'(0) \phi + \mathcal{O}(\phi^2)$.
The first term does not modify GR while $f'(0)$ in the second term can be absorbed to $\alpha$, hence we assume $f(\phi) = \phi$.

The action shown in Eq.~\eqref{action} should be treated as an \textit{effective theory}: a truncation of some more fundamental or complete theory (such as superstring theory) to second order in a curvature expansion.
It is only valid when the quadratic curvature term is smaller than the linear curvature term (i.e. $\alpha \phi R_\mrm{GB}^2/(\kappa R) < 1$).
In order to obtain the small coupling condition that can be applied to a vacuum spacetime as well, it would be convenient to extend this to $\alpha \phi K^2/(\kappa K) = \alpha \phi K/\kappa \lesssim 1$, where $K^2$ is defined as $K^2 \equiv C_{\mu \nu \rho \sigma} C^{\mu \nu \rho \sigma}$ with $C_{\mu \nu \rho \sigma}$ representing the Weyl tensor.
The correction relative to GR becomes the largest at the smallest length scale of the system.
For a LMXB, this corresponds to the horizon radius of a BH, where $\phi \sim \alpha/m_{1}^2$~\cite{Yunes:2011we} and $K \sim m_{1}^{-2}$ with $m_1$ representing the mass of the BH.
Therefore, the condition that the effective theory remains to be valid is given by $\alpha \phi K/\kappa \sim \alpha^2 /(\kappa m_1^4) <1$.
Hence, throughout this letter, we apply the small coupling approximation where  $\zeta \equiv \alpha^2/(\kappa m_1^4) < 1$.
This is a reasonable approximation given that GR has passed many tests in the weak field regime.
We only keep to linear order in $\zeta$, which means that we only take $\phi$ up to $\mathcal{O}(\zeta^{1/2})$. 
We can safely neglect the higher-order scalar field kinetic terms that appear in e.g. Ref.~\cite{amendola}.
The small-coupling approximation guarantees that the field equations to remain quadratic order in derivatives, which leads to the system being stable.

In this theory, an object acquires a scalar charge proportional to the volume integral of the Gauss-Bonnet invariant $R_\mrm{GB}^2$~\cite{quadratic}.
Since $R_\mrm{GB}^2$ is a topological invariant, the volume integral vanishes for any simply connected, asymptotically flat spacetime.
This means that the scalar charge vanishes for a star, including a NS, and only a BH can acquire a non-vanishing scalar charge. 
To be precise, the scalar charge of a star does not exactly vanish since the universe is not simply connected due to other BHs.
However, its effect can be safely neglected as long as its mass length scale is much smaller than the distance to its nearest BH.

\textit{Orbital decay rate of a LMXB:}
Let us consider a circular LMXB consisting of a BH with mass $m_1$, a main sequence star with mass $m_2$ and radius $R_2$ and having a binary separation $a$.
In Ref.~\cite{quadratic}, we derived the energy flux $\dot{E}$ due to GW and scalar radiation, where the origin of the leading correction relative to GR is the scalar dipole radiation and the order is of ``-1 post-Newtonian (PN)''.
By using the relation $\dot{E}=\Omega \dot{L}$ that holds under a circular-orbit binary, with $\Omega = \sqrt{m/a^3}$ denoting the orbital angular velocity, we derive the decay rate of the orbital angular momentum $L=\mu \sqrt{m a}$ as~\footnote{There are conservative corrections that would modify the expression of $\Omega$ from GR, but since their contributions are much smaller compared to the one from -1PN scalar radiation~\cite{quadratic}, we neglect them in this letter.}
\be
\dot{L} = \dot{L}_\mrm{GR} \left( 1+A v^a \right)\,.
\label{L}
\ee
Here, $\mu$ is the reduced mass, $m$ denotes the total mass, $v = \sqrt{m/a}$ is the typical velocity of the binary constituents, the parameters $A$ and $a$ are given by~\footnote{This can be obtained by setting $q_2=0$ in Eq.~(132) of Ref.~\cite{quadratic}. Notice that we have slightly changed the definition of $\zeta$ from Ref.~\cite{quadratic}.}
\be
A=\frac{5}{96} \zeta\,, \quad a=-2\,,
\label{A}
\ee
and $\dot{L}_\mrm{GR} = -(32/5) \eta^2 m v^7$ with $\eta \equiv m_1 m_2/m^2$ representing the symmetric mass ratio.

Eqs.~\eqref{L} and~\eqref{A} can be safely used for LMXBs whenever the small coupling approximation is valid.

The orbital period $P=2 \pi \sqrt{a^3/m}$ changes with time due to GW radiation and mass loss from the system by the stellar wind~\footnote{We do not consider the effects of magnetic braking and the evolution of the companion star on $\dot{P}$ since they are expected to be smaller than the observed bound~\cite{johannsen1,johannsen2}. }. 
The decay rate of the orbital period $P$ for a LMXB can be estimated as~\cite{zaglauer,psaltisBD}
\be
\frac{\dot{P}}{P} = 3 \left( \frac{n}{D} \right) \frac{\dot{L}}{L}\,,
\label{dotP}
\ee
where $n$ and $D$ are given in Ref.~\cite{psaltisBD} in terms of the following parameters: 
the mass ratio $q\equiv m_1/m_2$, the mass transfer efficiency $\beta_{\dot{m}} \equiv -\dot{m}_1/\dot{m}_2$, the specific angular momentum $j_w$ carried away by the stellar wind in units of $2 \pi a^2/P$ and the adiabatic index $\xi_\mrm{ad} \equiv d \ln R_2/d \ln m_2$.
We assume that $\beta_{\dot{m}}$ and $j_w$ both take values from 0 to 1, while we fix $\xi_\mrm{ad}$ as $\xi_\mrm{ad} = 0.8$~\cite{johannsen1,johannsen2}.

\begin{table*}[t]
\renewcommand{\arraystretch}{1.2}
\caption{\label{table} The observed parameters (the primary mass $m_1$, the mass ratio $q$, the orbital period $P$, the orbital decay rate $\dot{P}$, the eccentricity $e$ and the inclination $i$) of LMXBs A0620-00 and XTE J1118+480.
}
\begin{center}
\begin{tabular}{c||c|c|c|c|c|c}  
System & $m_1 \ (M_\odot)$ & $q$ & $P \ (\mrm{hr})$ & $\dot{P} \ (\mrm{s/s})$ & $e$ & $i$ \\ \hline\hline
A0620-00 & $6.6 \pm 0.25$~\cite{Cantrell:2010vh} & $17\pm 1$~\cite{johannsen1} & 7.8~\cite{johannsen1}  & $(1.66 \pm 2.64) \times 10^{-11}$~\cite{johannsen1} & - & $51^{\circ} \pm 0.9^{\circ}$~\cite{Cantrell:2010vh} \\ 
XTE J1118+480 & $8.30^{+0.28}_{-0.14}$~\cite{gonzalez} &  $37.0 \pm 12.3$~\cite{gonzalez} & 4.08~\cite{gonzalez} & $-(5.8 \pm 2.1) \times 10^{-11}$~\cite{gonzalez} & $< 0.0067 $~\cite{gonzalez} & $68^{\circ} \pm 2^{\circ}$~\cite{gelino} \\
\end{tabular}
\end{center}
\end{table*}

\textit{A new constraint on sGB gravity from A0620-00:}
Now, we derive the upper bound on $\zeta$ from the system A0620-00 whose parameters are summarized in the first row of Table~\ref{table}. 
By using the observed bound on $\dot{P}$
and Eqs.~\eqref{L} and~\eqref{dotP}, we obtain the 2-$\sigma$ upper bound on $\zeta$ as $\zeta < 7.3 \times 10^{-2}$.
(Here, we maximized $\zeta$ over (i) $\beta_{\dot{m}}$ and $j_w$ in the range 0 to 1, and (ii) $\dot{P}$ within 2-$\sigma$ range.) 
This leads to the 2-$\sigma$ constraint on $\sqrt{|\alpha|}$ as 
\be
\sqrt{|\alpha|} < 1.9 \times 10^{5} \ \mrm{cm}\,.
\label{alphaA}
\ee
This new constraint is more than six orders of magnitude stronger than the solar system bound~\cite{amendola}.
Eq.~\eqref{alphaA} is the main result of this letter.
Even if we restrict our attention to EDGB theory, it is still slightly stronger than the bound from GRO J0422+32.

One might think that the constraint $\zeta < 7.3 \times 10^{-2}$ validates our assumption of small coupling approximation only marginally.
However, we cannot completely rule out the possibility that the small coupling approximation is violated because at this stage, it would be difficult to estimate the amount of contribution coming from the higher curvature terms that we have truncated.
If the small coupling approximation is violated, it is either the case where (i) higher curvature contribution is larger than $\mathcal{O}(\zeta)$ one, or that (ii) these two give comparable contributions.
For the former, there would be much larger corrections relative to GR, and hence Eq.~\eqref{alphaA} still holds as a \textit{conservative} bound.
For the latter, if the signs of the two contributions are opposite, $\mathcal{O}(\zeta)$ correction would be reduced by the higher curvature one and the bound becomes weaker than Eq.~\eqref{alphaA}.
However, notice that the bound on $\sqrt{|\alpha |}$ scales with $A^{1/4}$ (see Eqs.~\eqref{L} and~\eqref{A}).
This means that even if the correction term in Eq.~\eqref{L} is reduced by a factor of 10, the bound on $\sqrt{|\alpha |}$ is only weakened by a factor of 2.
Moreover, this is a rather fine-tuned case compared to the case (i) or the one where  the small coupling approximation is valid.
For these reasons, we expect that the order of magnitude of the conservative bound 
$\sqrt{|\alpha |} \lesssim \mathcal{O}(10^5)$cm should still hold even if the small coupling approximation is not valid.

\textit{Explaining the excess in the orbital decay rate of XTE J1118+480 in sGB gravity:}
For  a LMXB XTE J1118+480 (whose observed parameters are summarized in the second row of Table~\ref{table}), the observed orbital decay rate is more than 10 times larger than the upper bound predicted in GR as $|\dot{P}| < 2.0 \times 10^{-12}\mrm{s/s}$.
The cause of this excess is most likely to be of astrophysical origin.
However, we want to point out that this excess can also be explained by the additional scalar radiation in sGB gravity.
In order to explain this observed value, we found that $\sqrt{|\alpha|}$ is required to be in the range $1.1 \times 10^{5} \mrm{cm} < \sqrt{|\alpha|} < 3.4 \times 10^{5} \mrm{cm}$.
(Here, we have both minimized and maximized $\sqrt{|\alpha|}$ over $\beta_{\dot{m}}$, $j_w$, $\dot{P}$ and $q$, where the range of the allowed value for $q$ is shown in Table~\ref{table}. We have included $q$ since its determination accuracy is not so high compared to other parameters.)
Combining this with Eq.~\eqref{alphaA}, we conclude that if $\sqrt{|\alpha|}$ lies in the range
\be
1.1 \times 10^{5} \ \mrm{cm} < \sqrt{|\alpha|} < 1.9 \times 10^{5} \ \mrm{cm}\,,
\label{range}
\ee
the excess can be explained by the scalar radiation in sGB gravity.
If we fix $q=37$, the only difference in Eq.~\eqref{range} is that the lower bound changes to $1.2 \times 10^{5}$cm. 

The allowed range in Eq.~\eqref{range} is rather small, but the point here is that at this stage, we cannot completely rule out the possibility of the excess due to the additional scalar radiation.
Below, we propose several ways to further pin down the value of $\alpha$ or that can completely rule out this possibility.

\textit{Ways to test sGB gravity as an explanation for the excess:}
It would be interesting to test Eq.~\eqref{range} with future observations.
One possibility is to look at the orbital decay rates of other BH-LMXBs since if the excess is caused by the scalar radiation, this effect should be universal to all similar systems.
Alternatively, if the measurement accuracy of $\dot{P}$ in A0620-00 could be improved, it might be possible to probe the range found in this letter. 
However, in order to realize this goal, the observation accuracy must be improved by about an order of magnitude, which is quite challenging.
In principle, if $\ddot{P}$ can be measured in the future, it may be possible to determine the cause of this excess (whether it is caused by the scalar radiation in sGB gravity, by the magnetic braking or by some other effect), since the correction in the energy flux should appear at different PN orders. 
However, this can only be done when the measurement accuracy of $\dot{P}$ improves considerably and this approach is much harder than the one previously mentioned.

Another possibility is that future GW interferometers may be able to distinguish it.
In Ref.~\cite{quadratic}, we found that the second-generation ground-based detectors such as adv. LIGO~\cite{adv-LIGO1} are likely to place a constraint $\sqrt{|\alpha|} < 4 \times 10^5$cm with a BH binary of $(6+12)M_\odot$ and a signal-to-noise ratio (SNR) of 20.
Unfortunately, this is not sufficient to probe the values shown in Eq.~\eqref{range} (and also the constraint is beyond the validity of the small coupling approximation).
For the Einstein Telescope (ET)~\cite{punturo}, which is roughly 10 times more sensitive than adv. LIGO, 
the constraint is still slightly larger than the upper bound in Eq.~\eqref{range} provided the upper bound on $\sqrt{|\alpha|}$ scales as (SNR)$^{1/4}$.
Next, let us consider space-borne GW interferometers.
By performing (sky-averaged) Fisher analyses~\cite{cutlerflanagan} explained in Refs.~\cite{yagiLISA,yagi:brane}, we found that ELISA~\cite{elisa} can give the constraint $\sqrt{|\alpha|} < 1.0 \times 10^5$cm  for a $(10+10^5)M_\odot$ circular, spin-aligned BH binary at 1Gpc for 1yr observation.
Furthermore, DECIGO/BBO~\cite{setoDECIGO,phinneybbo} will be able to constrain $\sqrt{|\alpha|} < 1.4 \times 10^3$cm for a $(1.4+10)M_\odot$ NS/BH binary with $\mrm{SNR}=10$ for 1yr observation. 
As discussed in Ref.~\cite{yagi:brane}, DECIGO/BBO is expected to detect about $N \sim 10^5$ NS/BH binaries.
Following Ref.~\cite{yagiDECIGO}, we expect that these multiple-source detection would make the constraint of $\zeta$ stronger by roughly $\sqrt{N}$.
Since $\sqrt{|\alpha|} \propto \zeta^{1/4}$, we expect that the constraint on $\alpha$ can be improved to $\sqrt{|\alpha|} \lesssim 10^2$cm.
This is indeed three orders of magnitude stronger than the new constraint from A0620-00 (Eq.~\eqref{alphaA}).
We also expect that the results are almost the same for precessing binaries~\cite{yagiLISA,yagiDECIGO}.
If we include the eccentricity $e$, it would correlate strongly with $\zeta$ since the correction terms from sGB and eccentricity both become greater for larger separation.
However, for the same reason discussed in Ref.~\cite{yagiLISA}, we have checked that this degeneracy is expected to be solved by adding prior information ($e^2 >0$) to the eccentricity.

\textit{Discussions:}
It seems difficult to explain the excess in the orbital decay rate of XTE J1118+480 in the context of most alternative theories of gravity, including massless and massive Brans-Dicke~\cite{brans,alsing}, dynamical Chern-Simons (CS)~\cite{alexanderyunes}, (non-linear) massive gravity theories~\cite{derham1,derham2}, Einstein-Aether~\cite{jacobson,foster} and Ho$\check{\mrm{r}}$ava-Lifshitz~\cite{horava,blas} gravities, since most of these theories have already been constrained rather strongly from binary pulsars and other tests such as solar system experiments~\cite{cassini} and galactic observations~\cite{sjors}.
In particular, even though it has not been constrained strongly yet, it would be difficult for the theories mentioned above to satisfy the requirements from the binary pulsar tests and also explain the excess in the LMXB. 
This is because the former observations are more precise than the latter and are known to agree with GR up to $\mathcal{O}(10^{-2})$. 
The interesting point about sGB gravity is that the scalar radiation appears only from binaries that contain BHs.
This means that sGB gravity cannot be constrained from binary pulsar observations.
Hence, it can explain the excess within the current bound.

In this letter, we have used the values of binary parameters that have been determined \textit{in GR}.
This can be justified as follows:
For sGB gravity, in Ref.~\cite{quadratic}, we discussed that there should be two effects that give non-dissipative corrections to GR, (i) the scalar forces and (ii) the deformation in the metric.
Since the secondary stars do not have scalar charges, we can forget about the former effect. 
For the latter, there would be a 2PN correction, as found by Yunes and Stein~\cite{Yunes:2011we}, of $\mathcal{O} \left( \zeta v^4 \right) \lesssim 10^{-11}$, which is way below the measurement error in mass.
Therefore, we can safely neglect the effect of modifications to the gravitational theory on the measured values of binary parameters.

In Ref.~\cite{quadratic}, we also found the dissipative correction in a compact binary system in CS gravity.
Therefore, it would be interesting to obtain new constraints on this theory from current binary observations.
We expect that the double-pulsar binary~\cite{yuneshughes} imposes a stronger constraint than LMXBs since the observation accuracy is higher for the former case.
In order to investigate this goal, we first need to derive the \textit{conservative} correction in this theory, which comes from the deformation in the metric at the quadratic order in spins and a dipole-dipole force that acts on each of the binary component.
We expect that the double-pulsar binary constraint should be stronger than the current solar system bound~\cite{yacine}.
This work is currently in progress.

It would be interesting to extend our formalism developed in Ref.~\cite{quadratic} for eccentric binaries and for non-vanishing scalar potential, and apply them to various binary systems.
These issues are left for future work.

The author thanks Takahiro Tanaka, Nicol\'as Yunes and Leo Stein for fruitful discussions and valuable comments.
The author also thanks Jonathan White for carefully reading this manuscript and giving us useful advice.
The author is supported by a Grant-in-Aid through the Japan Society for the Promotion of Science (JSPS) No. $22\cdot 900$.
This work is also supported in part by the Grant-in-Aid for the Global
COE Program ``The Next Generation of Physics, Spun from Universality and
Emergence'' from the Ministry of Education, Culture, Sports, Science and
Technology (MEXT) of Japan.


\bibliography{ref}
\end{document}